\documentclass[11pt,a4paper]{article}
\usepackage[utf8]{inputenc}

\usepackage{physics}
\usepackage{amsmath,amsfonts,amssymb}
\usepackage[pdftex]{graphicx}
\usepackage{color}

\newcommand{\ba}{\begin{aligned}}%
\newcommand{\ea}{\end{aligned}}%
\newcommand{\gst}{g}
\newcommand{\gpb}{\gamma}
\newcommand{\Gpt}{|\gpb_{tt}|}
\newcommand{\gpx}{\gpb_{\L\L}}
\newcommand{\gpz}{\gpb_{zz}}
\newcommand{\gos}{G}
\renewcommand{\L}{1}
\newcommand{\T}{2}
\newcommand{\pd}{\partial}

\newcommand{\Sp}{S_{\mathrm{D7}}}
\newcommand{\Lp}{\mathcal{L}_{\mathrm{D7}}}
\newcommand{\Tp}{T_{\mathrm{D7}}}
\newcommand{\SR}{\tilde{S}_{\mathrm{D7}}}

\newcommand{\cc}{\expval{\bar{\psi}\psi}}

\newcommand{\ap}{\mathcal{A}}

\newcommand{\rmS}{\mathrm{S}}
\newcommand{\rmU}{\mathrm{U}}
\newcommand{\calN}{\mathcal{N}}
\newcommand{\calA}{\mathcal{A}}

\renewcommand{\o}{\omega}
\newcommand{\s}{\sigma}

\begin{document}
\baselineskip=15.5pt
\pagestyle{plain}
\setcounter{page}{1}


\begin{flushright}
{\tt}
\end{flushright}

\vskip 2cm

\begin{center}
	{\Large \bf Mechanism for Negative Differential Conductivity\\ in Holographic Conductors}
\vskip 1cm

{\bf Shuta Ishigaki, Shin Nakamura} \\
\vskip 0.5cm
{\it  Department of Physics, Chuo University, \\
1-13-27 Kasuga, Bunkyo-ku, Tokyo  112-8551, Japan \\}
{\tt  E-mail: ishigaki@phys.chuo-u.ac.jp, nakamura@phys.chuo-u.ac.jp} \\
\medskip
\end{center}

\vskip1cm

\begin{center}
{\bf Abstract}
\end{center}
\medskip
We clarify the mechanism for negative differential conductivity in holographic conductors. Negative differential conductivity is a phenomenon in which the electric field decreases with the increase of the current. This phenomenon is widely observed in strongly correlated insulators, and it has been known that some models of AdS/CFT correspondence (holographic conductors) reproduce this behaviour.
We study the mechanism for negative differential conductivity in holographic conductors by analyzing the lifetime of the bound states of the charge carriers.
We find that when the system exhibits negative differential conductivity, the lifetime of the bound states grows as the electric field increases.
This suggests that the negative differential conductivity in this system is realized by the suppression of the ionization of the bound states that supplies the free carriers.

\newpage

\tableofcontents

\section{Introduction}
Negative differential conductivity (NDC) is a phenomenon where the electric field acting on the system decreases as the current increases.
Electric devices which exhibit NDC play an important role in construction of switching circuits and oscillator circuits.
NDC has been observed in a wide range of strongly-correlated electron systems \cite{HAoki2014}. However, we still lack a complete understanding of the mechanism for NDC on the basis of the microscopic theory.\footnote{The NDC of electric devices such as Esaki diode has been explained by the tunneling effect at the p-n junction. However, the NDC we consider is a non-ballistic transport of charged particles in bulk materials.}
A crucial reason for the difficulty is that we need to deal with physics far from equilibrium.

When we have a current along an electric field, the system is out of equilibrium owing to the Joule heating.
The system realizes a nonequilibrium steady state (NESS) when the work given by the external electric field and the dissipation into the heat bath are in balance.
In order to study the mechanism of NDC, we need to deal with NESSs in the nonlinear regime.

Recently, the AdS/CFT correspondence \cite{Maldacena1998,SGubser1998,EWitten1998} has been applied to analysis of NDC \cite{SNakamura2010Prog,SNakamura2012PRL,Ali-Akbari:2013hba,Matsumoto:2018ukk,Imaizumi:2019byu}.
The AdS/CFT correspondence states that a strongly-coupled quantum gauge theory is equivalent to a higher-dimensional classical gravity theory.
This enables us to compute the expectation values of physical quantities even out of equilibrium.
In \cite{SNakamura2010Prog,SNakamura2012PRL,Ali-Akbari:2013hba,Matsumoto:2018ukk,Imaizumi:2019byu}, the D3-D7 model \cite{AKarchEKatz2002,AKarchAOBannon2006} is employed as a gravity dual of a $(3+1)$-dimensional many-body system of charged particles interacting with a thermal reservoir.%
\footnote{The D3-D5 model for \(2+1\)-dimensional systems is also considered in \cite{Ali-Akbari:2013hba}.}
It has been found that this model shows NDC at low temperatures.
However, the mechanism for the NDC in this system has not been understood.

It is known that the system has bound states of a positively charged particle and a negatively charged particle \cite{Erdmenger:2007cm}. 
If we make an analogy with QCD, the bound states correspond to mesons, or bound states of quark and anti-quark. However, it would be more appropriate to make an analogy with strongly-correlated electron systems when our aim is investigation of nonlinear conductivity in materials. In this context, the bound state corresponds to an exciton which is a bound state of an electron and a hole.

In this paper, we analyze the lifetime of the bound states to find the mechanism of the NDC.
We employ the D3-D7 model with vanishing charge density.
The system is an insulator at sufficiently low temperatures.
If we apply a large enough electric field to the system, we can break the insulation. This is because the positively-charged particles (which we call positive carriers for short) and the negatively-charged particles (which we call negative carriers) will be created by the Schwinger effect.\footnote{This corresponds to the Landau-Zener effect in solid state physics.}
 
However, if the positive carriers and the negative carriers form neutral bound states, they do not contribute to the DC charge transport.
This implies that the formation of the bound states decreases the conductivity.
With this in mind, we study the relationship between the lifetime of the bound states and the electric field, systematically.
In our system, we have three possible origins of the charge carriers which contribute to the conductivity:
thermal excitation from the vacuum, pair creation from the vacuum induced by the external electric field, and ionization of the already-existing bound states by the electric field.
We will clarify which process is the main contribution to the realization of the NDC.


We find a counter-intuitive behaviour of the bound states 
when the system exhibits NDC: the lifetime of the bound states grows as the electric field increases.
We also find that the conductivity of the system approaches zero when the lifetime of the bound states vanishes when the system shows NDC.
This means that the charge carriers in the NDC regime are mainly supplied by the decay process of the bound states.
This explains the NDC well:
the larger the electric field, the longer the lifetime of the bound states, then the number density of charge carriers decreases and the current density decreases.

This paper is organized as follows.
In Sec.~\ref{sec:background}, we briefly review the D3-D7 model and how NDC is realized there \cite{SNakamura2010Prog, SNakamura2012PRL, AKarchAOBannon2006}.
In Sec.~\ref{sec:perturbation}, we compute the lifetime of the bound state in the D3-D7 model when the NDC is realized.
We show the electric-field dependence of the lifetime.
In Sec~\ref{sec:discussion}, we present discussion and conclusions.
Some related topics are given in Appendix \ref{apdx:scalar} and Appendix \ref{apdx:ingoing-wave}.
A possible relationship between our results and AC conductivity (optical conductivity) is given in Appendix \ref{apdx:ac-cond}.

\section{Setup}\label{sec:background}
We study a NESS with a constant electric current along an constant external electric field, by using the AdS/CFT correspondence.
The system is consisting of charged particles immersed in a heat bath.
We consider the case of vanishing charge density, where the number of the positively charged particles and that of the negatively charged particles are equal.
We assume the case that the system is homogeneous and steady.
In this section, we review how the NDC is realized in the D3-D7 model \cite{SNakamura2010Prog}.

We employ the D3-D7 model as the gravity dual \cite{AKarchEKatz2002,AKarchAOBannon2006}.
This model shares common features with strongly-correlated insulators. The system in the field-theory side is a $(3+1)$-dimensional system. The model has positive charge carriers and negative charge carriers. The interaction among (infinitely massive) charges are Coulomb type interaction at zero temperature. 
The system is an insulator at low temperature, whereas we have finite conductivity at high temperatures. Since our aim is qualitative explanation of NDC and clarification of its basic mechanism, we employ the D3-D7 model in this paper.

In this model, the heat bath is realized as an 5-dimensional asymptotically AdS-Schwarzschild black hole geometry times $\rmS^5$.
The spacetime metric is given by
\begin{equation}
	\dd s^2
	=\gst_{MN} \dd X^M\dd X^N
	=\frac{1}{z^2}\left[
	-\frac{(1-z^4/z_h^4)^2}{1+z^4/z_h^4}\dd t^2+(1+z^4/z_h^4)\dd\vec{x}^2
	\right]
	+\frac{\dd z^2}{z^2}
	+ \dd\Omega^2_5.
\end{equation}
We have set the AdS radius to 1.
$z=z_h$ is the location of the black hole horizon.
The Hawking temperature is given by $T=\frac{\sqrt{2}}{\pi z_h}$.
$\dd\Omega^2_5$ denotes the metric of the unit 5-sphere.

The system of charged particles is given as an D7-brane whose action is%
\footnote{
	We set $(2\pi\alpha')=1$.
}
\begin{equation}
	\Sp=-\Tp\int\dd^8\xi\sqrt{-\det(\gpb+F)},~~~
	\gpb_{ab}=\gst_{MN}\pdv{X^M}{\xi^a}\pdv{X^N}{\xi^b},
\end{equation}
where $\Tp$ is the tension of the D7-brane,
$\xi^a$ are the worldvolume coordinates,
$\gpb_{ab}$ is the induced metric and $F_{ab}=\pd_a A_b-\pd_b A_a$ is the $\rmU(1)$ field strength%
\footnote{
	We write $M,N$ as indices of the spacetime coordinates which run from 0 to 9,
	whereas $a,b$ are indices of the worldvolume coordinates which run from 0 to 7.
}.
In the D3-D7 model, we assume that the D7-brane are wrapped on the $\rmS^3$ part of the $\rmS^5$.
The metric of $\rmS^5$ can be written as
\( \dd\Omega^2_5 = \dd\theta^2+\sin^2\theta\dd\psi^2+\cos^2\theta\dd\Omega^2_3,\)
where \(0\leq \theta \leq \pi/2\) and \( 0 \leq \psi < 2\pi\) are the $\rmS^2$ coordinates.
We employ a static gauge $\xi^a=(t,\vec{x},z,\Omega^3)$, then $\theta$ and $\psi$ are dynamical fields.
We assume that $\theta$ depends only on $z$, and $\psi$ is set to zero by virtue of the symmetry.

We apply an electric field in the $x^\L$ direction.
The \(x^{\L}\)-component of the $\rmU(1)$ gauge field is $A_\L=-Et+a_\L(z)$ with an appropriate choice of the gauge.
By solving the equation of motion for $a_\L(z)$, we obtain
\begin{equation}
	J = \frac{\calN a_\L'(z)\Gpt\cos^3\theta
	}{\sqrt{\Gpt\gpx\gpz-(\gpz E^2-\Gpt a_\L'(z)^2)}},
\end{equation}
where $\calN\equiv\Tp(2\pi)^2$.
\(J\) is an integration constant which gives the expectation value of the current density in the \(x^{\L}\) direction.
The value of $J$ is determined as follows.
We find
\(
	a_{1}'(z)^2 = \frac{\gpx}{\Gpt}\frac{h(z)}{k(z)}\frac{J^2}{\calN^2},
\)
where
\(
	h(z)\equiv \Gpt\gpx-E^2,
\) \(
	k(z)\equiv \Gpt\gpx^3\cos^6\theta(z) - \gpx J^2/\calN^2.
\)
\(J\) is determined by requesting that \(h(z)\) and \(k(z)\) go across zero simultaneously at a same location, say, $z=z_*$: \( h(z_*)=k(z_*)=0\) \cite{AKarchAOBannon2006}.
From these conditions, \(z_*\) can be written as
\( z_*^2/z_h^2 = \sqrt{1+\left(E/\pi^2T^2\right)^2}-E/\pi^2T^2. \)
The conductivity \(\sigma_{\text{DC}}\equiv J/E\) is given by
\begin{equation}
	\sigma_{\text{DC}}= \calN \pi T \sqrt{\sqrt{1+\left(\frac{E}{\pi^2T^2}\right)^2}\cos^6\theta(z_*)},
	\label{eq:dc_conductivity}
\end{equation}
which shows a nonlinear conductivity \cite{AKarchAOBannon2006}.
The field $\theta(z)$ is related to the mass of the charged particles.
To obtain the conductivity for a specific finite mass theory, we have to solve $\theta(z)$ so that the mass of the charge carriers fixed at a designed value.
(See Appendix \ref{apdx:scalar}.)

\begin{figure}
	\centering
	\includegraphics[width=0.5\linewidth]{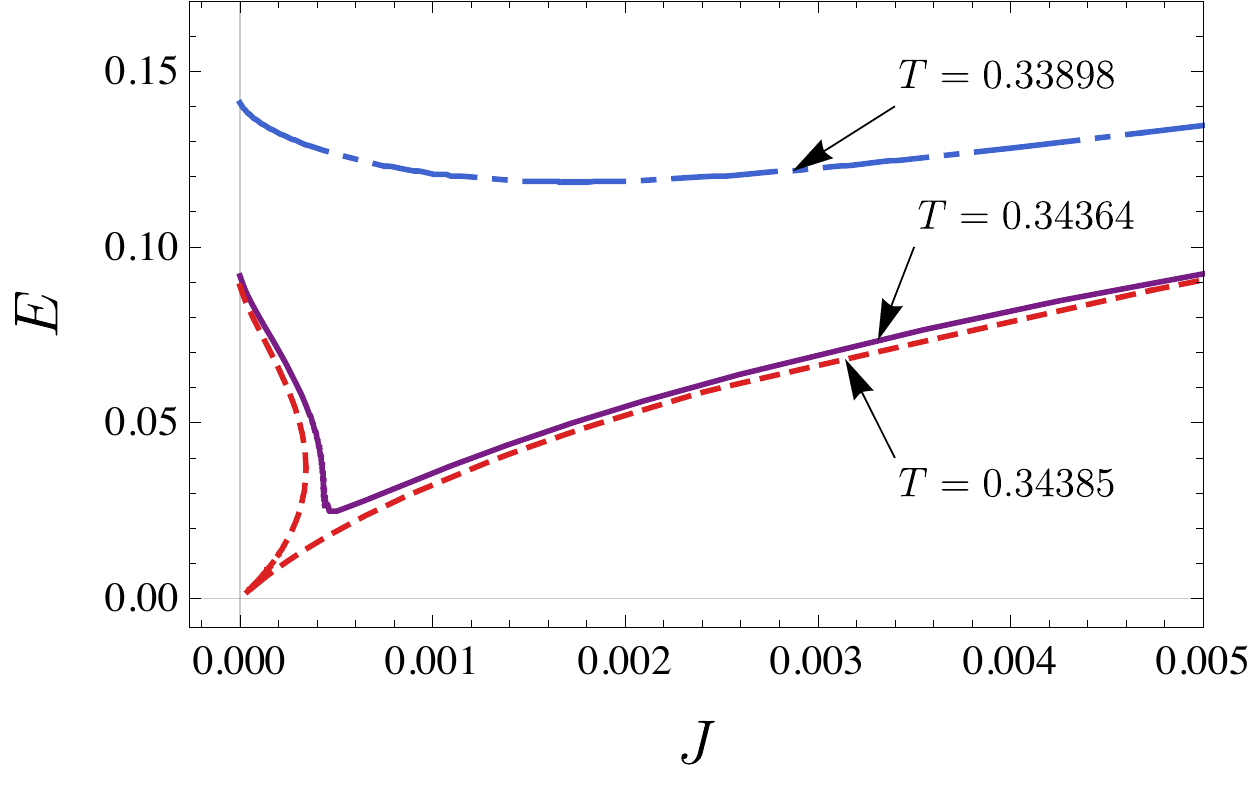}
	\caption{
		$J$-$E$ characteristics at various temperatures:
		from top to bottom $T=0.33898, 0.34364, 0.34385$ respectively.
		We have set $m_q=1$ for simplicity.
	}
	\label{fig:JE_wide}
\end{figure}
The $J$-$E$ characteristics in our system at various temperatures are given in Fig.\,\ref{fig:JE_wide}.
The system is an insulator, but the insulation is broken by applying a strong electric field.
In the small $J$ region, we can see NDC: $\partial J/\partial E < 0$ \cite{SNakamura2010Prog,SNakamura2012PRL}.%
\footnote{
	It is known that a system which has such a NDC often shows instability called as filamentary instability \cite{schoell}.
	There is a possibility that our system also shows the instability.
	It is interesting to investigate the instability however we don't consider the instability since we focus the homogeneous system in this paper.
}
We have set $m_q=1$ and $\calN=1$ for simplicity.

\section{Quasi-normal modes}\label{sec:perturbation}
Our goal is to study the behavior of the lifetime of the bound states of positive carrier and negative carrier when NDC is realized.
In the framework of the AdS/CFT correspondence, the dynamical modes are realized as the normalizable modes in the gravity dual.
The bound states mentioned above correspond to the normalizable modes on the D7-brane.%
\footnote{In terms of hadron physics, these bound states may be refereed to as mesons or bound states of quarks. For the correspondence between these states and the normalizable modes on the probe D-brane, see for example, Ref.~\cite{Erdmenger:2007cm}. }
Since the D7-barne intersects with the horizon of the black hole, the normalizable modes are quasi-normal modes (QNMs) with a finite lifetime.
A QNM has a complex-valued frequency, \(\omega = \omega_R + i \omega_I\) which realizes a damped oscillation.
\(\omega_R\) gives the energy of the bound state and \(-\omega_I\) gives the decay width of the bound state for our case.
The aim of this section is to analyze the behavior of \(\omega_R\) and \(\omega_I\) as a function of the external electric field \(E\).


\subsection{Setup}
\label{Setup_QNM}
A QNM is realized as a perturbation field on a background configuration.
In this study, we consider a perturbation of the normalizable mode of the transverse vector field for simplicity, since it is decoupled from all other perturbation fields at the linear order of the equation of motion.
Behavior of the spectral function of \(A_{\T}\) has been studied in \cite{JMas2009}.
We analyze the behavior of the QNM of the perturbation field of \(A_{\T}\) in detail and clarify its dependence on \(E\) in the NDC region.


The ansatz for the gauge fields is
\begin{equation}
	A_a\dd\xi^a \equiv (-E t + a_\L(z))\dd x^\L + \calA(z,t)\dd x^\T,
\end{equation}
where $\calA(z,t)$ is the perturbation field of the transverse vector mode.
We have set the transverse direction to the \(x^{\T}\) direction without loss of generality.
We assume the perturbation field is independent of the spatial coordinates $\vec{x}$.

The equations of motion for the
gauge field are written as
\begin{equation}
	\pd_b [\sqrt{-\det(\gpb+F)}\gpb^{ac}F_{cd}\gos^{db}]=0,
	\label{eq:gauge_fields_EoM}
\end{equation}
where $\gos_{ab}$ is the \textit{open string metric}:
\(
\gos_{ab}\equiv \gpb_{ab}-F_{ac}\gpb^{cd}F_{db}.
\)
$\gos^{ab}$ is defined by $\gos^{ac}\gos_{cb}={\delta^a}_b$.
We consider the Fourier transform of the gauge field,
$\tilde{\calA}(z,\omega)=\int_{-\infty}^{\infty} \dd t  \exp(i\omega t) \ap(z,t)$.
Then the equation of motion is written as
\begin{equation}
	\ba
	\pd_z& [\sqrt{-\det(\gpb+F)}\gpb^{\T\T}
	(\gos^{zz}\ap'(z)-i\o\gos^{tz}\ap(z))]\\
	&-i\o\sqrt{-\det(\gpb+F)}\gpb^{\T\T}\gos^{tz}\ap'(z)
	-\o^2\sqrt{-\det(\gpb+F)}\gpb^{\T\T}\gos^{tt}\ap(z)=0,
	\label{eq:EoM_vector_transverse}
	\ea
\end{equation}
where we simply wrote $\tilde{\ap}(z,\o)$ as $\ap(z)$.
In order to solve this equation properly, we should impose the ingoing wave boundary condition at \(z=z_*\).
Let us perform the Frobenius expansion at \(z=z_*\): \(\ap(z)=(1-z/z_*)^{i\lambda}\ap_{\text{reg}}(z)\).
Though there are two distinct choices of \(\lambda\) allowed from the equation of motion, the ingoing wave boundary condition is achieved by choosing \(\lambda=0\).
(See Appendix \ref{apdx:ingoing-wave}.)

Since \(\gos^{zz}\) vanishes at \(z=z_*\), we obtain
\begin{equation}
	\frac{\ap'(z_*)}{\ap(z_*)}
	=\left.
	\frac{
		i\o(\sqrt{-\det(\gpb+F)}\gpb^{\T\T}\gos^{tz})'/\sqrt{-\det(\gpb+F)}\gpb^{\T\T}
		+\o^2\gos^{tt}
	}{
		(\gos^{zz})'-2i\o\gos^{tz}
	}
	\right|_{z=z_*}.
\end{equation}
We also need to impose the vanishing condition at the boundary: $\ap(0)=0$ at $z=0$.
This condition ensures that the solution is a resonance state without an external source.
We solve Eq.\,(\ref{eq:EoM_vector_transverse}) under these conditions by the shooting method.\footnote{
	It has been found that the numerical errors in the shooting method are large in the D3-D7 model when $|\o_I|$ is greater than the order of the temperature \cite{MKaminski/JErdmenger2010}.
	However, we discuss only the small \(|\o_I|\) region in this paper. 
}
Note that the location of the pole of QNM depends on $E, J, T$, since Eq.~(\ref{eq:EoM_vector_transverse}) depends on them.

\subsection{The behavior of the bound states}\label{sec:QNMresult}

\begin{figure}[htb]
	\centering
	\includegraphics[width=0.6\linewidth]{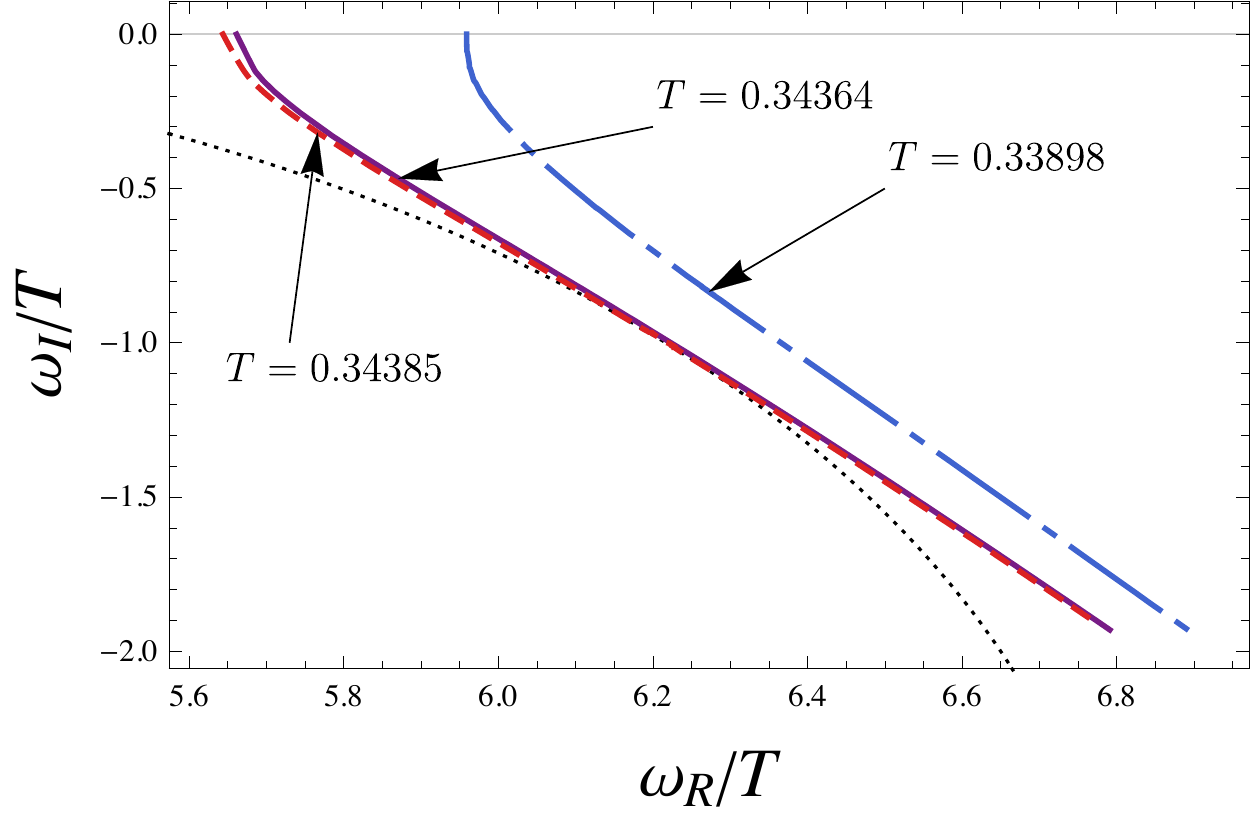}
	\caption{
		Trajectories of the poles at fixed temperatures when we vary \(E\) and $J$ along the corresponding $J$-$E$ curve in Fig.\,\ref{fig:JE_wide}.
		The dotted line is a reference line at \(E=0\) for various temperatures.
	}\label{fig:Omegaplane}
\end{figure}

\begin{figure}[htb]
	\centering
	\hspace{-45pt}
	\includegraphics[width=0.65\linewidth]{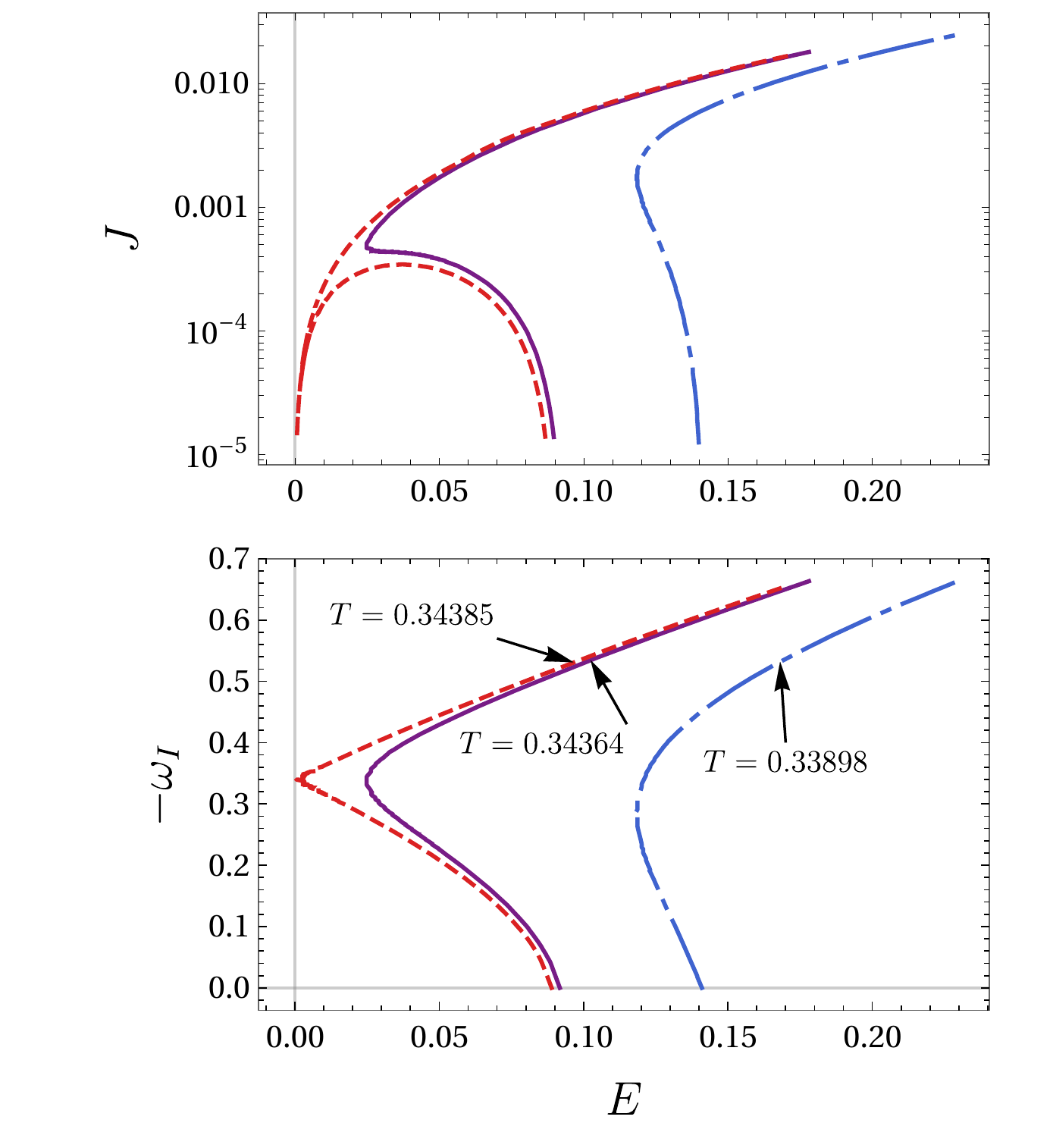}
	\caption{
		(top) $E$ dependence of $J$,
		(bottom) $-\o_I$ as a function of $E$
		at various temperatures.
	}
	\label{fig:E_vs_width}
\end{figure}

\begin{figure}[htbp]
	\centering
	\hspace{30pt}
	\includegraphics[width=0.7\linewidth]{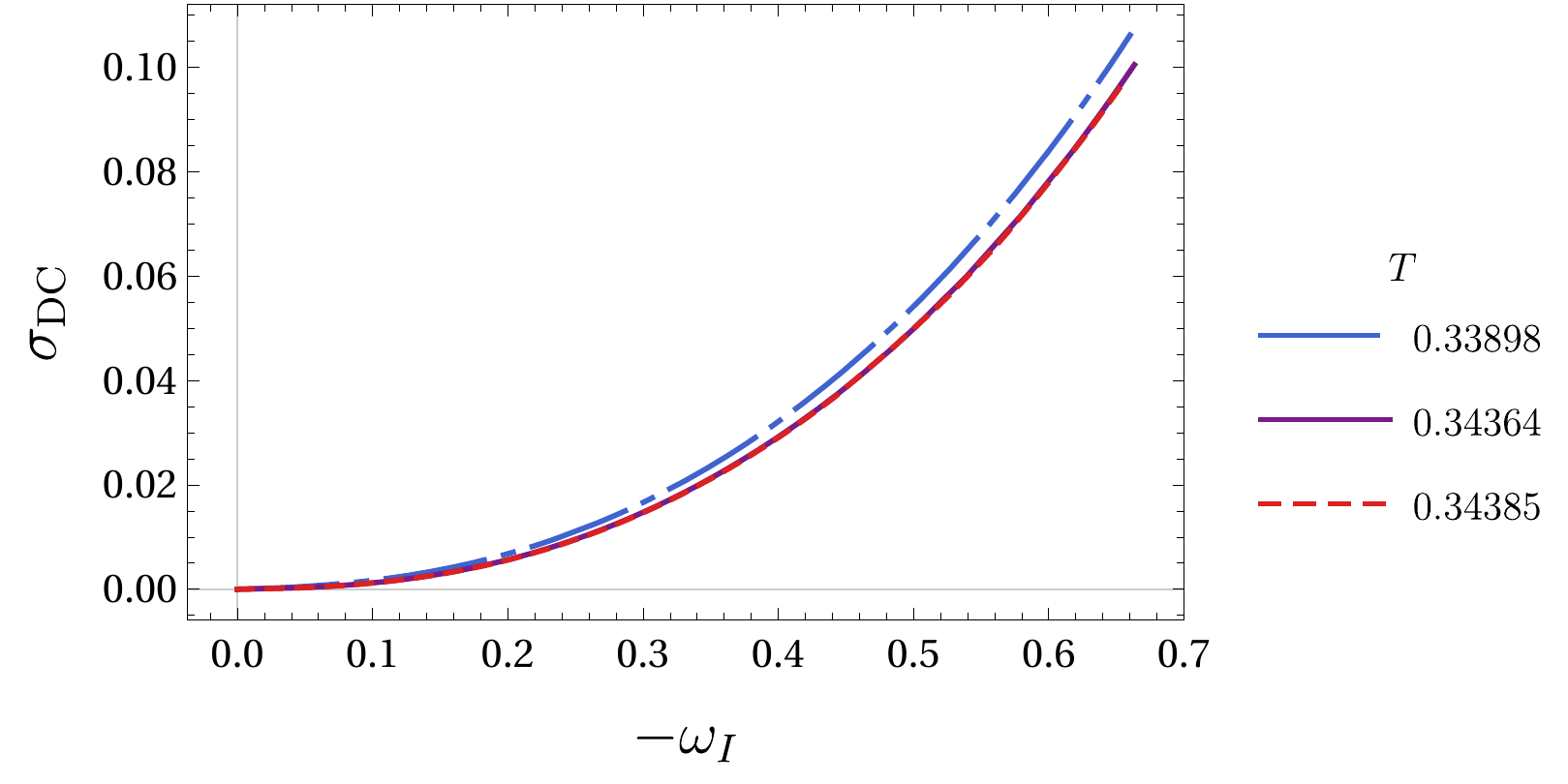}
	\caption{
		DC conductivity $\s_{\mathrm{DC}}$ vs. decay width $-\o_I$
		at various temperatures.
		The lower two lines are almost overlapping.
	}
	\label{fig:Conductivity_vs_width}
\end{figure}

We have computed $\omega_{R}$ and $\omega_{I}$ of the first excited mode of the transverse vector field as functions of $E$ at various temperatures. 
In Fig.\,\ref{fig:Omegaplane}, Fig.\,\ref{fig:E_vs_width} and Fig.\,\ref{fig:Conductivity_vs_width},
the dash-dot line, the solid line, and the broken line represent the data at $T=0.33898, 0.34364$ and $0.34385$, respectively.
The \(J\)-\(E\) characteristics at the same temperatures are given in Fig.\,\ref{fig:JE_wide}.



The location of the pole of the QNM is given in Fig.\,\ref{fig:Omegaplane}.
The pole moves along the curve at each temperature when we vary $E$ and $J$ along the corresponding $J$-$E$ curve in Fig.\,\ref{fig:JE_wide}.
The dotted line is a reference which indicates the location of the pole at \(E=0\) for various temperatures.
We have checked that the reference line asymptotes to the analytical result \(\omega/T=2\pi(1-i)\) \cite{RCMyersAOStarinetsRWThomson2007}
at the high temperature limit.
Note that the broken line at \(T=0.33898\) touches the reference line of \(E=0\).
This is because the \(J\)-\(E\) characteristics at this temperature reaches \(E=0\) as shown in Fig.\,\ref{fig:JE_wide}.
The lifetime of the bound state is inversely proportional to the imaginary part ($-\omega_{I}$) of the corresponding QNM: the behavior of the lifetime of the bound state is directly read from the analysis of the QNM given in Section~\ref{Setup_QNM}.

Let us see the details of the behavior of the lifetime of the bound states under the presence of \(E\).
The top of Fig.\,\ref{fig:E_vs_width} shows the $J$-$E$ characteristics
and the bottom shows the relationship between the decay width $-\o_I$ and the electric field $E$.
Note that $J$ and $-\omega_{I}$ are multi-valued functions of $E$.%
\footnote{Note that NDC in strongly-correlated insulators is observed in current-controlled experiments~\cite{HAoki2014}: $E$ and $-\o_I$ are single-valued function of $J$, for $T=0.33898$ and $T=0.34364$. We find $E$ and $-\o_I$ are still multi-valued function of $J$ for $T=0.34385$ which indicates the presence of the current-driven nonequilibrium phase transition discovered in \cite{SNakamura2012PRL}. The origin of this multi-valued behavior is not clear, but its clarification is beyond the scope of the present paper and we leave it for future work.} The branches of the smaller value of $-\omega_{I}$ correspond to the branches of smaller $J$. 
For the solid line and the dash-dot line, the smaller-$J$ branches always show NDC. 
We find that $\pd(-\o_I)/\pd E<0$ for these branches.
The fact that \(\pd(-\o_I)/\pd E\) turns into negative when the system shows NDC is the main result of our analysis.
Let us find the physical implication of this result.
It is helpful to see Fig.\,\ref{fig:Conductivity_vs_width} for this purpose.

Fig.\,\ref{fig:Conductivity_vs_width} shows the relationship between the decay width $-\o_I$ and the DC conductivity $\s_{\mathrm{DC}}$ given in Eq.\,(\ref{eq:dc_conductivity}).
$-\o_I$ is a monotonically increasing function of $\s_{\mathrm{DC}}$ irrespective of the temperature.
In particular, $\s_{\mathrm{DC}}$ reaches zero when $-\o_I=0$: we do not have electric conduction if the bound state is completely stable.
This shows that the electric conduction in the vicinity of $\s_{\mathrm{DC}}=0$ in the present system is owing to the charge carriers supplied by the ionization of the bound states.



Now the physical implication of our result becomes clear.
In the vicinity of \(\s_{\mathrm{DC}}=0\), the charge carriers are supplied by the ionization of the bound states.
Let us focus on the \(J\)-\(E\) characteristics \textit{in the vicinity of} \(\sigma_{DC}=0\) in Fig.~\ref{fig:E_vs_width}.
We should refer to the data where \(E\) remains finite when \(J\) approaches zero there.
One finds that the data in the vicinity of \(\s_{\mathrm{DC}}=0\) show NDC, and $\pd(-\o_I)/\pd E<0$.
These observations lead us to the conclusion that \textit{the origin of the NDC in the present system is suppression of the ionization of the bound states which provides the charge carriers, by the increase of the electric field.}



\section{Discussion and Conclusions}\label{sec:discussion}


In this paper, we analyzed the lifetime of the bound states by using AdS/CFT correspondence to reveal the mechanism of NDC.
We found that the lifetime of the bound states becomes longer as the electric field increases when the system shows NDC. We also found that the charge carriers in our system is mainly produced by the ionization process of the bound states in the parameter region where NDC is realized.
These leads us to the conclusion that the origin of the NDC in the present system is suppression of the ionization of the bound states which provides the charge carriers, by the increase of the electric field.

Let us make a few comments.
Our statement is that when we have NDC, $\pd(-\o_I)/\pd E<0$. However we do {\em not} claim that whenever $\pd(-\o_I)/\pd E<0$ is realized the system always shows NDC. 
For example, the lower branch of the broken line at $T=0.34385$ in the bottom of Fig.\,\ref{fig:E_vs_width} shows PDC even though $\pd(-\o_I)/\pd E<0$.

Our statement in this paper is only for NDC, and we do not attempt to make any concrete statement for PDC.
However, we can make the following discussions.
Since our system is neutral, the possible origins of the charge carriers are the following three processes: i) thermal excitation of carriers from the vacuum, ii) the Schwinger effect induced by the electric field that creates pairs of positive and negative charge carriers, and iii) the ionization of the (already-existing) bound states by the electric field.  
What we have found is that when NDC is realized, the main origin of the charge carriers is iii). However, the above three processes can contribute to the electric conduction in general. For example, when we raise the temperature, the process i) will not be negligible.
This implies that the mechanism of NDC we proposed above will not work in the high-temperature region. Indeed,  
the system shows the $J$-$E$ characteristics given by $J\approx \calN E^{3/2}$, which gives PDC, at the high-temperature limit.
Therefore, we should consider not only the process iii) but also others for PDC.

We would also like to make a comment on possible connection to experiments.
The bound state in the present system is corresponding to the excitons in the electron-hole systems in the context of solid state physics.
It is interesting to see the electric-field dependence of the lifetime of the excitons in the materials which show NDC, experimentally. The lifetime of excitons can be experimentally detected by observing the optical conductivity (AC conductivity). We present a short review on the relationship between the lifetime of the bound states and the optical conductivity in Appendix~\ref{ACcond}.
It has been discussed that excitons may from the exciton Bose-Einstein condensation at low temperatures \cite{exciton_bec}.
It is worth while studying how the exciton Bose-Einstein condensation affects the non-linear conductivity of the materials.

\subsection*{Acknowledgements}
We would like to thank
Y.~Fukazawa, T.~Hayata,
H.~Hoshino, S.~Kinoshita and R.~Yoshii
for helpful discussions and comments.
The work of S.~N.\ was supported in part by JSPS KAKENHI Grants No. JP16H00810, No. JP19K03659, No. JP19H05821, and the Chuo University Personal Research Grant.
The work of S.~I.\ was supported by the Research Assistant Fellowship of Chuo University.

\appendix

\section{Equation of motion for scalar field}\label{apdx:scalar}
We derive the equation of motion for the scaler field \(\theta(z)\).
Let us consider a Legendre transform of DBI action called Routhian:
\begin{equation}
\ba
\SR\equiv&\Sp - \int\dd za_{\L}'(z)\pdv{\Lp}{a_{\L}'(z)}\\
=&-\calN\int\dd z\dd t
\sqrt{\frac{\gpz}{\Gpt\gpx}}
\sqrt{
	(\Gpt\gpx-E^2)
	(\Gpt\gpx^3\cos^6\theta(z) - \gpx J^2/\calN^2)}.
\ea
\label{eq: Routhian}
\end{equation}
We obtain the equation of motion from Eq.\,(\ref{eq: Routhian}) as:
\begin{equation}
\pdv{}{z}\left(
\theta'(z)\sqrt{\frac{h(z)k(z)}{\Gpt\gpx\gpz}}
\right)
-3\sin\theta\cos^5\theta
\sqrt{\Gpt\gpx^5\gpz \frac{h(z)}{k(z)}}=0,
\label{eq:EoM-theta}
\end{equation}
where \(h(z)\) and \(k(z)\) are defined in Sec.~\ref{sec:background}.
The boundary condition for the scaler field at \(z=z_*\) is given by taking the limit \(z\to z_*\) of Eq.\,(\ref{eq:EoM-theta}).
The condition is written as
\begin{equation}
	\theta'(z_*)=\frac{b-\sqrt{b^2+c^2}}{z_* c},
	\label{eq:theta-prime}
\end{equation}
where
\[
\ba
	b&\equiv(3z_*^8+2z_*^4 z_h^4+3z_h^8),&
	c&\equiv3(z_*^4-z_h^4)(z_*^4+z_h^4)\tan\theta(z_*).&
\ea
\]
Eq.\,(\ref{eq:theta-prime}) means that \(\theta'(z_*)\) is given in terms of \(\theta(z_*)\), hence we obtain a unique solution once we determine the value of \(\theta(z_*)\).
\(\theta(z)\) in the vicinity of $z=0$ is given by
\begin{equation}
\theta(z)= m_q z + \left(\frac{\cc}{2\calN}+\frac{m_q^3}{6}\right)z^3+\cdots,
\end{equation}
where $m_q$ is the mass of the charge carrier%
\footnote{
	The mass corresponds to the band gap in the condensed matter physics.
}
 and $\cc$ is the chiral condensate.
 Therefore \(m_q\) is determined if we assign the value of \(\theta(z_*)\).
 We keep \(m_q\) fixed at a designed value by choosing the value of \(\theta(z_*)\) appropriately at each given value of \(E\).

\section{Ingoing wave boundary condition at the effective horizon}\label{apdx:ingoing-wave}

We impose the ingoing wave boundary condition at \(z=z_*\) for the perturbation field \(\ap\) when we study QNMs.
This is because the position of \(z=z_*\) on the worldvolume of the D7-brane plays a role of horizon for the perturbation field: we call \(z=z_*\) as effective horizon.
Suppose that \(\ap\) is given as follows in the vicinity of \(z=z_*\),
\begin{equation}
	\ap(z,\o)=(1-z/z_*)^{-i\lambda}(\ap_{(0)} + \order{1-z/z_*}).
\end{equation}
\(\lambda\) is determined by Eq.\,(\ref{eq:EoM_vector_transverse}).
Note that \(\gos^{zz}\) goes to zero whereas \(\gos^{tz}\) and \(\gos^{tt}\) do not vanish at \(z=z_*\).
Then we find that \(\lambda\) must satisfy
\begin{equation}
	\lambda\left(
	\lambda+2\o\left.\frac{\gos^{tz}}{(\gos^{zz})'}\right|_{z_*}
	\right)= 0.
\end{equation}
We obtain
\begin{equation}
	\lambda=
	\begin{cases}
	\lambda_-=0,\\
	\lambda_+=-2\o \left.\gos^{tz}/(\gos^{zz})'\right|_{z_*}.
	\end{cases}
\end{equation}
In \cite{JMas2009}, it is argued that \(\lambda_-\) is corresponding to  the ingoing wave boundary condition in view of the consistency in the limit of \(E\rightarrow 0\).

We obtain the same conclusion by using conformally flat coordinates as follows.
The equation of motion of the perturbation field is described by using the open string metric \(\gos_{ab}\), which has off-diagonal components \(\gos_{tz}=\gos_{zt}\) in our setup.
The open string metric can be diagonalized as
\begin{equation}
	\gos_{ab}\dd\xi^a\dd\xi^b=
	|\gos_{tt}|\left[
	-\left(
		\dd t - \frac{\gos_{tz}}{|\gos_{tt}|}\dd z
	\right)^2
	+\frac{1}{|\gos_{tt}|}\left( \gos_{zz} + \frac{\gos_{tz}^2}{|\gos_{tt}|}\right) \dd z^2
	\right] + \cdots.
\end{equation}
We consider a conformally flat coordinates
\renewcommand{\tt}{\tilde{t}}
\newcommand{\zt}{\tilde{z}}
\begin{equation}
	\tt = t - \int_{0}^{z} \frac{\gos_{tz}}{|\gos_{tt}|}\dd z',~~~
	\zt = -\int_{0}^{z} \sqrt{
	\frac{1}{|\gos_{tt}|}\left(\gos_{zz} + \frac{\gos_{tz}^2}{|\gos_{tt}|}\right)
	} \dd z',
\end{equation}
so that we have
\begin{equation}
	\gos_{ab}\dd\xi^a\dd\xi^b=
	|\gos_{tt}|\left( -\dd\tt^2 + \dd\zt^2 \right) + \cdots.
\end{equation}
In the vicinity of $z=z_*$, we have
\begin{equation}
	\tt =  t - \left.\frac{\gos_{tz}}{|\gos_{tt}|'}\right|_{z_*}\log|1-z/z_*|,~~~
	\zt = - \left.\frac{\gos_{tz}}{|\gos_{tt}|'}\right|_{z_*}\log|1-z/z_*|.
\end{equation}
The perturbation field in the vicinity of \(z=z_*\) is now given by
\begin{equation}
	\ap_{(0)} \exp\left[-i\omega(\tt \pm \zt)\right],
\end{equation}
where the double sign corresponds to that of \(\lambda_{\pm}\) respectively.
Since the effective horizon is at \(\zt= + \infty\), the ingoing wave solution is \(\ap_{(0)} \exp\left[-i\omega(\tt - \zt)\right] = \ap_{(0)} \exp\left[-i\omega t\right]\) which agrees with the statement of \cite{JMas2009}.

\section{AC conductivity}\label{apdx:ac-cond}
\label{ACcond}

\begin{figure}[htbp]
	\centering
	\includegraphics[width=0.5\linewidth]{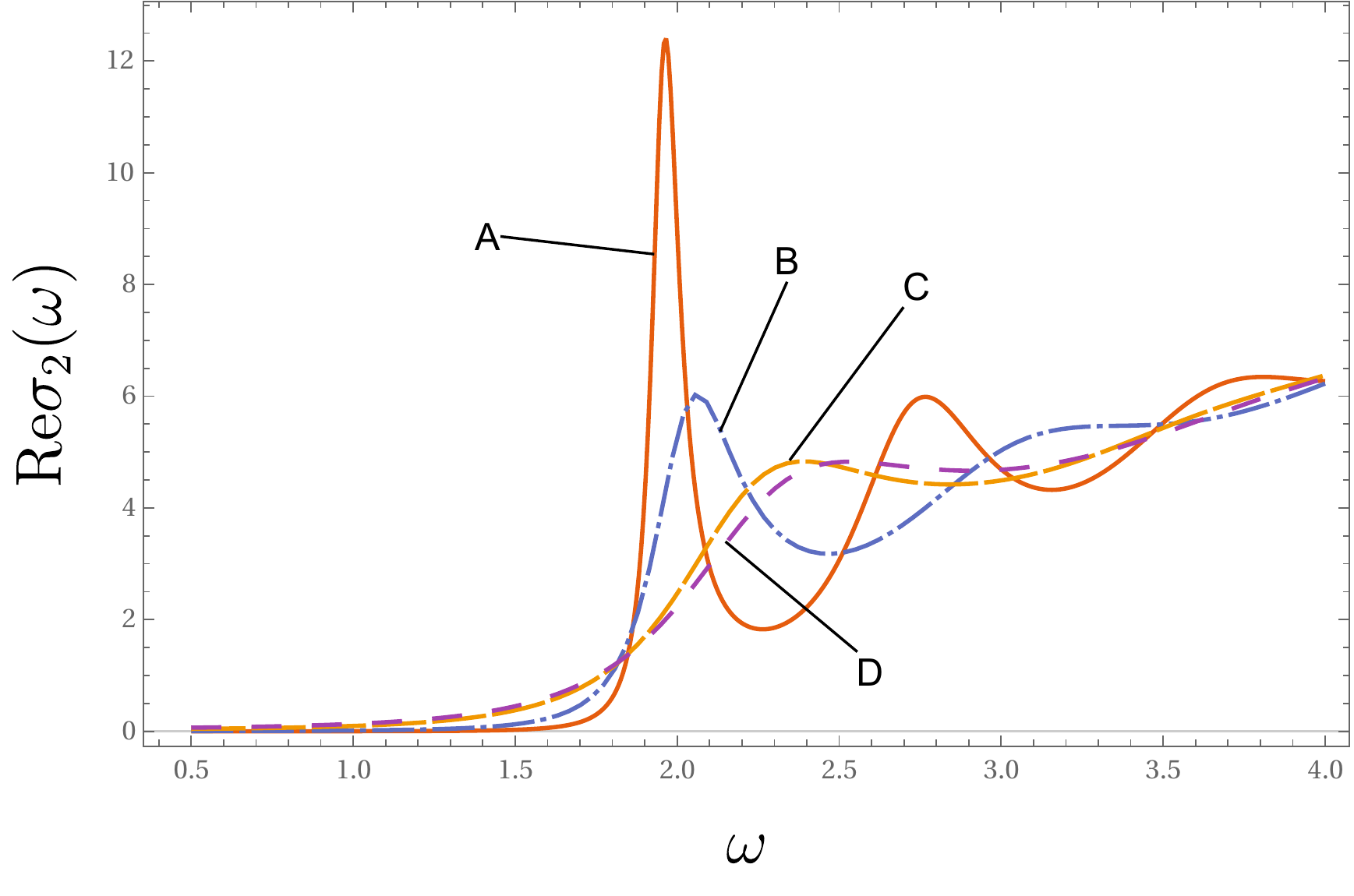}
	\includegraphics[width=0.25\linewidth]{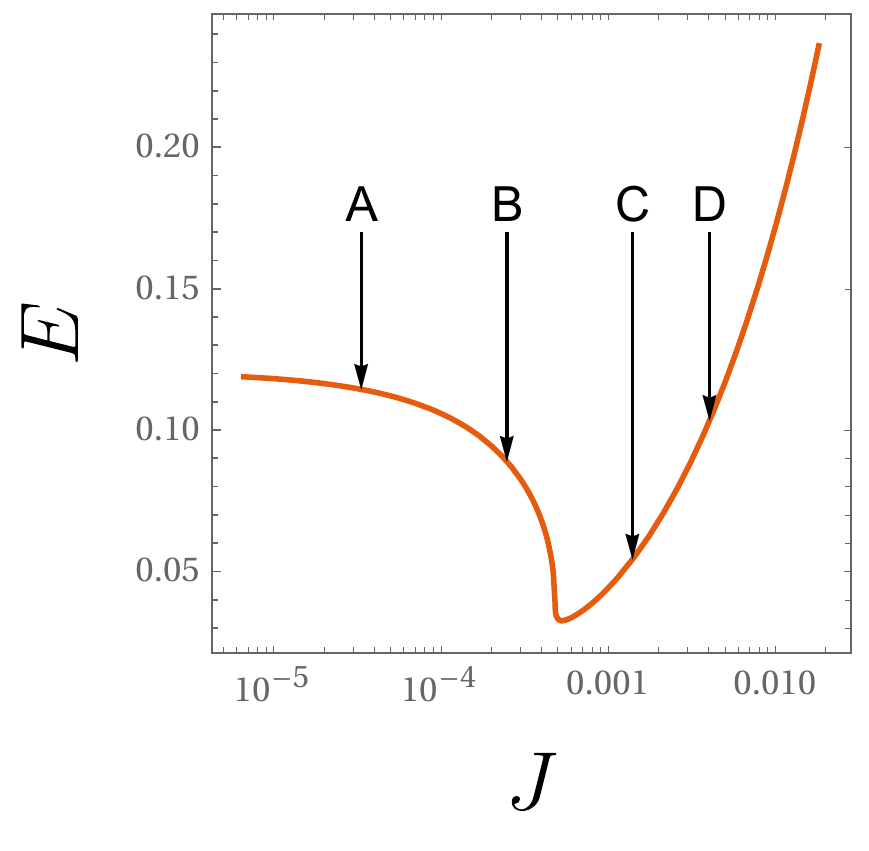}
	\caption{
		(Left) AC conductivity,
		(Right) \(J\)-\(E\) characteristic  at \(T=0.34364\).
		The value of \(J\) for each curve in the left figure are corresponding to those labeled by A to D on the \(J\)-\(E\) curve.
	}\label{fig:spec}
\end{figure}

We have studied the lifetime of the bound state via the imaginary part of the QNM-frequency.
Alternatively, one can read it from the full width at half maximum (FWHM) of peaks in the AC conductivity.
Since the AC conductivity is measurable in experiments, it is worth while reviewing how the lifetime comes into the AC conductivity.
Let us assume that the standard oscillator model called the Lorentz model \cite{wooten1972} can be applied to our system.
In the oscillator model, the dielectric function is given by
\begin{equation}
	\epsilon(\omega)
	=
	\epsilon_{\infty}
	+ \sum_{n}\left[
		\frac{
			- S_{n} \mathcal{E}_{n}^2 (\omega^2 - \mathcal{E}_{n}^2)
		}{
			(\omega^2 - \mathcal{E}_{n}^2)^2 + \omega^2 \gamma_{n}^2
		}
		+ i
		\frac{
			S_{n} \omega \gamma_{n} \mathcal{E}_{n}^2
		}{
			(\omega^2 - \mathcal{E}_{n}^2)^2 + \omega^2 \gamma_{n}^2
		}
	\right],
\end{equation}
where \(S_{n}\), \(\mathcal{E}_{n}\) and \(\gamma_{n}\) are the oscillator strength, the oscillator frequency and the damping factor of each oscillator labeled by \(n\).
\(\epsilon_{\infty}\) is the dielectric constant at high frequency.
The oscillator frequency and the damping factor are corresponding to the energy and the decay width of the bound states, respectively.
The AC conductivity is related to the dielectric function by \(\sigma(\omega) = -i \omega \epsilon(\omega)\).
The real part of the AC conductivity is expressed as
\begin{equation}
	\mathrm{Re} \sigma(\omega) = \sum_{n}
	\frac{
		S_{n} \omega^2 \gamma_{n} \omega_{n}^2
	}{
		(\omega^2 - \omega_{n}^2)^2 + \omega^2 \gamma_{n}^2
	}.
\end{equation}
\( \mathrm{Re}\sigma(\omega) \) has a peak at \(\omega=\omega_n\) when \(\gamma_n/\omega_n\) is sufficiently small.
\(\gamma_{n}\) is the FWHM of each peak centering at \(\omega=\omega_n\).

In our system, the real part of the AC conductivity%
\footnote{
	The spectrum function corresponding to the AC conductivity in this setup has been investigated in \cite{JMas2009}.
}
 has peaks as shown in Fig.\,\ref{fig:spec}.
We can see the peak located at \(\omega\approx 2\) broadens as \(J\) increases along the curve in the right figure of Fig.\,\ref{fig:spec}.
This behavior is consistent with our result shown in Fig.\,\ref{fig:E_vs_width} at \(T=0.34364\) that \(-\omega_I\) of the lowest mode increases when \(J\) grows.

\end{document}